\documentclass[a4paper,notitlepage,fleqn,12pt]{article}
\usepackage[tbtags]{amsmath}
\usepackage{amsfonts}
\usepackage{amsthm}
\usepackage{color}
\usepackage{graphicx,amssymb}
\usepackage{multirow}
\usepackage{rotating}
\usepackage{mathrsfs}
\usepackage{epstopdf}
\usepackage{enumerate}
\usepackage{pdflscape}
\usepackage{geometry}
\usepackage{natbib}
\usepackage{changepage}
\usepackage{dcolumn}
\usepackage{tabularx}
\usepackage{subfig}
\usepackage{mathabx}
\usepackage[mathscr]{euscript}
\usepackage{diagbox}

\usepackage{floatrow}
\usepackage{tikz}
\usetikzlibrary{graphs}
\usepackage{mathdots}
\usepackage{algorithm}
\usepackage{algorithmicx}
\usepackage{algpseudocode}
\usepackage{booktabs}

\newcolumntype{d}[1]{D{.}{.}{#1}}
\newcolumntype{Y}{>{\raggedleft\arraybackslash}X}
\newcolumntype{Z}{>{\centering\arraybackslash}X}

\theoremstyle{remark}

\DeclareMathOperator*{\argmin}{argmin}

\newcommand{\bm}[1]{\mbox{\boldmath{$#1$}}}
\numberwithin{equation}{section}

\usepackage[colorlinks,linkcolor=blue,citecolor=blue]{hyperref}

\begin{document}

\title{\vspace{-10mm}Effective sample size estimation based on concordance between $p$-value and posterior probability of the null hypothesis}

\author{
	Han Wang\thanks{College of Science, China Agricultural University, Beijing, China.}
	\and
	Yan Dora Zhang\thanks{Department of Statistics and Actuarial Science, School of Computing and Data Science, The University of Hong Kong, Hong Kong SAR, China.}
	\and
	Guosheng Yin\thanks{Department of Statistics and Actuarial Science, School of Computing and Data Science, The University of Hong Kong, Hong Kong SAR, China. *Corresponding author.}
}

\date{}
\maketitle

\begin{abstract}
	Estimating the effective sample size (ESS) of a prior distribution is an age-old yet pivotal challenge, with great implications for clinical trials and various biomedical applications. Although numerous endeavors have been dedicated to this pursuit, most of them neglect the likelihood context in which the prior is embedded, thereby considering all priors as ``beneficial". In the limited studies of addressing harmful priors, specifying a baseline prior remains an indispensable step. In this paper, by means of the elegant bridge between the $p$-value and the posterior probability of the null hypothesis, we propose a new ESS estimation method based on $p$-value in the framework of hypothesis testing, expanding the scope of existing ESS estimation methods in three key aspects: 
	(\text{i}) We address the specific likelihood context of the prior, enabling the possibility of negative ESS values in case of prior--likelihood disconcordance;
	(\text{ii}) By leveraging the well-established bridge between the frequentist and Bayesian configurations under noninformative priors, there is no need to specify a baseline prior which incurs another criticism of subjectivity;
	(\text{iii}) By incorporating ESS into the hypothesis testing framework, our $p$-value ESS estimation method transcends the conventional one-ESS-one-prior paradigm and accommodates one-ESS-multiple-priors paradigm, where the sole ESS may reflect the collaborative impact of multiple priors in diverse contexts. Through comprehensive simulation analyses, we demonstrate the superior performance of the $p$-value ESS estimation method in comparison with existing approaches. Furthermore, by applying this approach to an expression quantitative trait loci (eQTL) data analysis, we show the effectiveness of informative priors in uncovering gene eQTL loci.
\end{abstract}
\newpage

\section{Introduction}
Assessing the impact of priors on Bayesian statistical inference is crucial in clinical trial designs and other healthcase applications. Effective sample size (ESS) serves as an important quantitative metric reflecting the amount of information conveyed by the prior. 
Considerable effort has been devoted to impelling deeper understandings of ESS. One essential line is by virtue of baseline priors, while the ESS estimation in different methods are equipped with different metrics. In the work by \cite{clarke1996implications}, relative entropy is utilized to measure the distance between the target prior and the posterior under the baseline prior, and the ESS is defined as the sample size required in constructing the posterior to reach the minimum relative entropy. \cite{morita2008determining} adopted a similar approach. They formally defined an $\epsilon$-information prior as the baseline prior, possessing the same mean as the target prior but significantly inflated variance to guarantee its ``baseline" role. The metric is constructed by comparing the Fisher information of the posterior under the baseline prior and the curvature of the log density of the target prior. The ESS is then defined as the sample size needed to achieve the minimal information divergence. 
The baseline prior is also a necessary ingredient in the study by \cite{reimherr2021prior}, whereas the authors adpoted the posterior mean squared error (MSE) as a criterion to determine the optimal ESS. Differing from matching the target prior and the baseline prior corresponded posterior as in \cite{clarke1996implications} and \cite{morita2008determining}, \cite{reimherr2021prior} utilized the posterior--posterior matching based on the baseline and target prior respectively. The synergy of MSE together with posterior-posterior matching allows for a negative ESS, a stunning fact enforcing us to revisit a prior under different prior--likelihood contexts---the same prior may contribute positively to statistical inference in some cases, while diminishing the informative likelihood in others. Another key question is that, when it is challenging to identify the baseline prior, is it possible to circumvent specification of a baseline prior to fulfill the ESS estimation goal?

For the other line in which a baseline prior is not essential, \cite{neuenschwander2020predictively} introduced various ways to interpret the ESS. For instance, when the sample size $n$ explicitly appears in the updating rule from prior to posterior parameters, the ESS can be identified automatically. In many conjugate cases, the posterior mean is a weighted average of the prior mean and the sample mean, with weights proportional to the ESS and sample size $n$, respectively. Therefore, the ESS can be obtained by the form of posterior mean. Additionally, the ESS can be estimated by connecting the prior's variance to the variance of an estimator $Y_n$ for $\theta$ from a sample of size $n$. In this case, the ESS is defined as the sample size such that the two variances are equal. This variance-matching method is intuitive, especially in the case of a normal distribution. When a conjugate normal prior is assigned to the mean parameter $\mu$, the prior variance can be exactly regarded as a precision level where how many samples are required to reach. The work by \cite{malec2001closer}, \cite{pennello2007experience} and \cite{neuenschwander2010summarizing} also involved variance-ratio methods to derive the ESS. \cite{neuenschwander2020predictively} proposed a novel ESS estimation rule that integrates variance-ratio and information-based methods. In contrast to the work by \cite{morita2008determining}, where the divergence measurement is evaluated at the prior mean, the ESS defined by \cite{neuenschwander2020predictively} is the expectation of the ratio of prior information to Fisher information. For this class of ESS estimation, although there is no need to specify a baseline prior, it is apparent that the bilateral characteristics for a target prior under different prior--likelihood contexts cannot be addressed.

In this paper, we propose a novel approach to address the ESS estimation problem. The core idea draws inspiration from the connection between the $p$-value and posterior probability in the hypothesis testing framework proposed in \cite{dudley2002asymptotic}. Due to the well-established bridge between the frequentist configuration and Bayesian configuration under noninformative priors, it is not necessary to specify a baseline prior in the ESS estimation procedure. Additionally, by inserting a Bayesian prior into the hypothesis testing system, the discordance between the prior and the likelihood can be assessed. Thus, the $p$-value ESS esimation procedure embraces both prior--likelihood concordant and conflict settings, allowing for both positive and negative ESS.

The remaining part of the paper is structured as follows. In section 2, we first present an overall picture of the $p$-value ESS estimation method, and then delve into the details of its estimation methodology in four different settings. In section 3, numerous simulation studies are conducted to showcase the effectiveness and convenience of the $p$-value ESS estimation procedure under different prior--likelihood contexts. Section 4 presents a real data application, where the ESS is embedded into an expression quantitative trait loci (eQTL) analysis. We show that when a well-established prior is imposed, the enhanced power in eQTL detection is effectively demonstrated by the sample size gain estimated through the $p$-value ESS estimation procedure. In section 5, we conclude by addressing the advantages, limitations and further explorations.

\section{Methodology}

Based on the pioneering work of \cite{dudley2002asymptotic}, it has been established that for one-sided tests, the $p$-value is asymptotically equivalent to the posterior probability of the null hypothesis under noninformative priors \citep{shi2021reconnecting}. 
Considering the one-sample scenario, we are interested in the one-sided hypothesis test, say, $H_0: \theta\leq a$ versus $H_1: \theta> a$ based on sample $X_1,\ldots,X_n$. Assume that $f(x|\theta)$ denotes the density function of $X_1,\ldots,X_n$ parametrized by $\theta$. Now, if an informative prior $p(\theta)$ supports the null hypothesis, it's expected that the posterior probability of the null hypothesis increases because the prior leans towards the null hypothesis, i.e., this informative prior contributes to the loss of hypothesis testing power. On the other hand, back to the $p$-value side, it is well-known that when the sample size decreases, the $p$-value increases. Therefore, from the frequentist perspective, it is the sample size that drives power loss; whereas from the Bayesian perspective, it is the informative prior that drives power loss. When these two power values are sacrificed to the same extent, the ESS for the informative prior can be reflected by the change in sample size in the frequentist framework. 

In the context of two-sample scenarios, there is often a keen interest in comparing the magnitudes of parameters in the two groups, i.e., $H_0: \theta_1\leq \theta_2$ versus $H_0: \theta_1> \theta_2$, where we assume $f(x|\theta_1)$ and $f(x|\theta_2)$ are density functions of $X_1^{(1)},\ldots,X_n^{(1)}$ and $X_1^{(2)},\ldots, X_n^{(2)}$ respectively. Even if two priors $p_1(\theta_1)$ and $p_2(\theta_2)$ are imposed for $\theta_1$ and $\theta_2$ respectively, only one posterior probability is derived for inference. Similar to the ESS derivation principle elucidated in the one-sample scenario above, there is only one induced ESS that showcases the synergy of two priors. Apparently, in the context of two-sample  comparison, our prior assessment of the distance between two groups, whether optimistic or pessimistic, may inflate or deflate the posterior probability of the null hypothesis. Consequently, the deduced ESS encompasses both positive and negative values, in line with the optimistic and pessimistic roles of prior specification. In light of two-sample comparsion, our $p$-value ESS estimation procedure surpasses the previous one-ESS-one-prior regime, allowing for the one-ESS-multiple-priors occasion. Owing to the hypothesis testing framework, the $p$-value ESS estimator also manifests the promising or disheartening role of priors in distinguishing between the null and alternative hypothesis. 

Obviously, owing to the agency role of the $p$-value, there is no need to specify a baseline prior. Additionally, the prior--likelihood concordance or disconcordance can be reflected by different levels of posterior probability. Therefore, the $p$-value ESS estimation method allows for interpreting priors under different prior--likelihood deviation circumstances. 

\subsection{ESS of normal prior}\label{motivate_normal}
We first explore the ESS of a normal prior under the one-sample normal distribution case. Suppose that $X_i\sim N(\mu,\sigma^2)$ for $i=1,\ldots,n$, and 
our interest lies in the one-sided hypothesis test
$$H_0: \mu\leq 0\ \ \text{versus}\ \  H_1: \mu>0.$$
Denote $\bar{X}_n=\sum_{i=1}^n X_i/n$, and then the frequentist test statistic is $Z_n^F=\sqrt{n}\bar{X}_n/\sigma$. If a normal prior is imposed on the mean parameter $\mu$, i.e., $\mu\sim N(\delta,\sigma^2/m)$, the posterior distribution for $\mu$ is given by
$$\mu\sim N\left(\frac{\sum_{i=1}^n X_i+m\delta}{m+n},\frac{\sigma^2}{m+n}\right).$$ 

For simplicity, let $\tilde{\mu}$ and $\tilde{\sigma}^2$ represent the posterior mean and variance of $\mu$ respectively, and then the corresponding $Z$-statistic in the Bayesian setting is denoted as $Z_n^B=\tilde{\mu}/\tilde{\sigma}$. Without loss of generality, we assume the underlying truth $\mu=0$. When the prior has no deviation from the true value, i.e., $\delta=0$, $Z_n^B$ is observed to be a rescaled version of $Z_n^F$, with the degree of rescaling determined by the intensity of the prior, which is reflected by $m$. 

To account for all possible realizations, we introduce two metrics based on the two $Z$-statistics $Z_n^B$ and $Z_{\tilde{n}}^F$,
$$U_n^B=E(Z_n^{B})^2,\ \ \text{and}\ \  U_{\tilde{n}}^F=E(Z_{\tilde{n}}^F)^2,$$
where $\tilde{n}$ is the adjusted sample size in the frequentist scheme. Based on these two metrics, it is straightforward to define a distance $D(n,\tilde{n})=|U_n^B-U_{\tilde{n}}^F|$ to quantify the extent of testing power gain or loss for $Z_n^B$ compared with $Z_{\tilde{n}}^F$, and then we take $\tilde{n}^*=\argmin_{\tilde{n}}|D(n,\tilde{n})|$. When the underlying true distribution falls in the null hypothesis zone, this situation corresponds to a small value of $U_n^F$. The prior resulting in a smaller value of $U_n^B$ than $U_n^F$ is regarded as a ``good" prior, and under this situation, $\tilde{n}^*\leq n$. While the prior resulting in a larger value of $U_n^B$ may hinder us from making the correct decision in accepting the null hypothesis, we regard such priors as ``poor" priors, and under this situation, $\tilde{n}^*>n$. Therefore, when the data support $H_0$, we define $\textrm{ESS}=n-\tilde{n}^*$.

It is worth noting that by the properties of $Z$-score, $E(\sqrt{n}\bar{X}_n/\sigma)^2$ is always equal to 1 with respect to all $n$. To reflect the change of sample size, we vary $\tilde{n}$ in $Z_{\tilde{n}}^F$ while keeping the value of $\bar{X}_n$ intact, i.e., we define $Z_{\tilde{n}}^F=\sqrt{\tilde{n}}\bar{X}_n/\sigma$.

Based on the definition of $U_n^B$ and $U_{\tilde{n}}^F$, the explicit form of the ESS can be derived:
\begin{itemize}
	\item If there is no deviation, i.e., $\delta=0$, $D(n,\tilde{n}^*)=0$ when $\tilde{n}^*=n^2/(m+n)$. This leads to $\textrm{ESS}=n-\tilde{n}^*=mn/(m+n)$.
	\item If there exists deviation, i.e., $\delta\neq 0$, $D(n,\tilde{n}^*)=0$ when $\tilde{n}^*=m^2\delta^2n/(m+n)$. This leads to $\textrm{ESS}=n-\tilde{n}^*=mn(1-m\delta^2)/(m+n)$.
\end{itemize}	

\begin{figure}
	\centering         
	\includegraphics[width=1\textwidth]{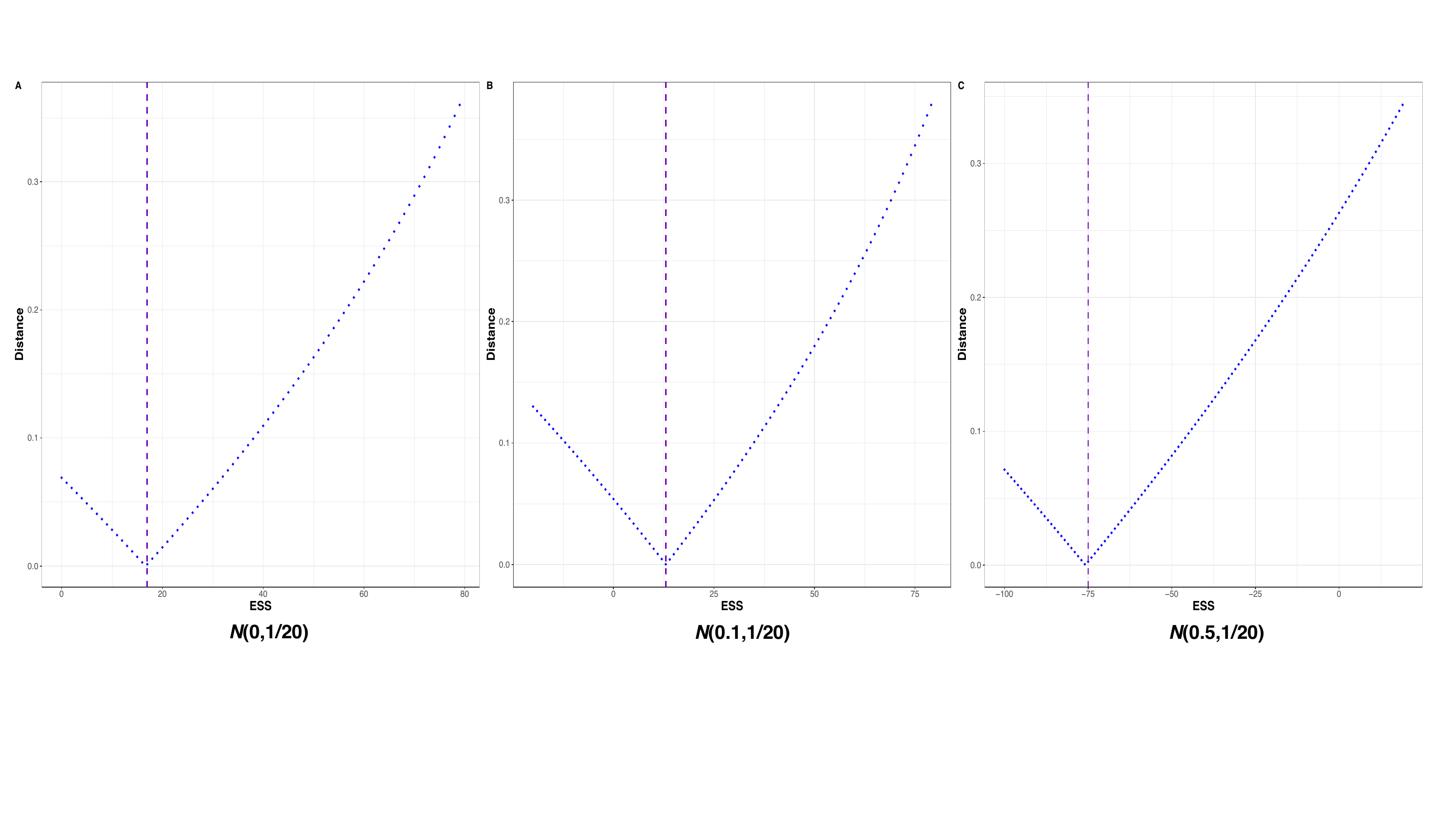}
	\caption{\textit{The distance $D(n,\tilde{n})=D(n,n-\textrm{ESS})$ versus ESS under three priors: $N(0,1/20)$, $N(0.1,1/20)$, and  $N(0.5,1/20)$. The minimum value of $D(n,n-\textrm{ESS})$ are obtained at $\textrm{ESS}=17, 13$ and $-79$  in (A), (B) and (C) repectively.}}   
	\label{fig1}
\end{figure}

\begin{figure}
	\centering         
	\includegraphics[width=1\textwidth]{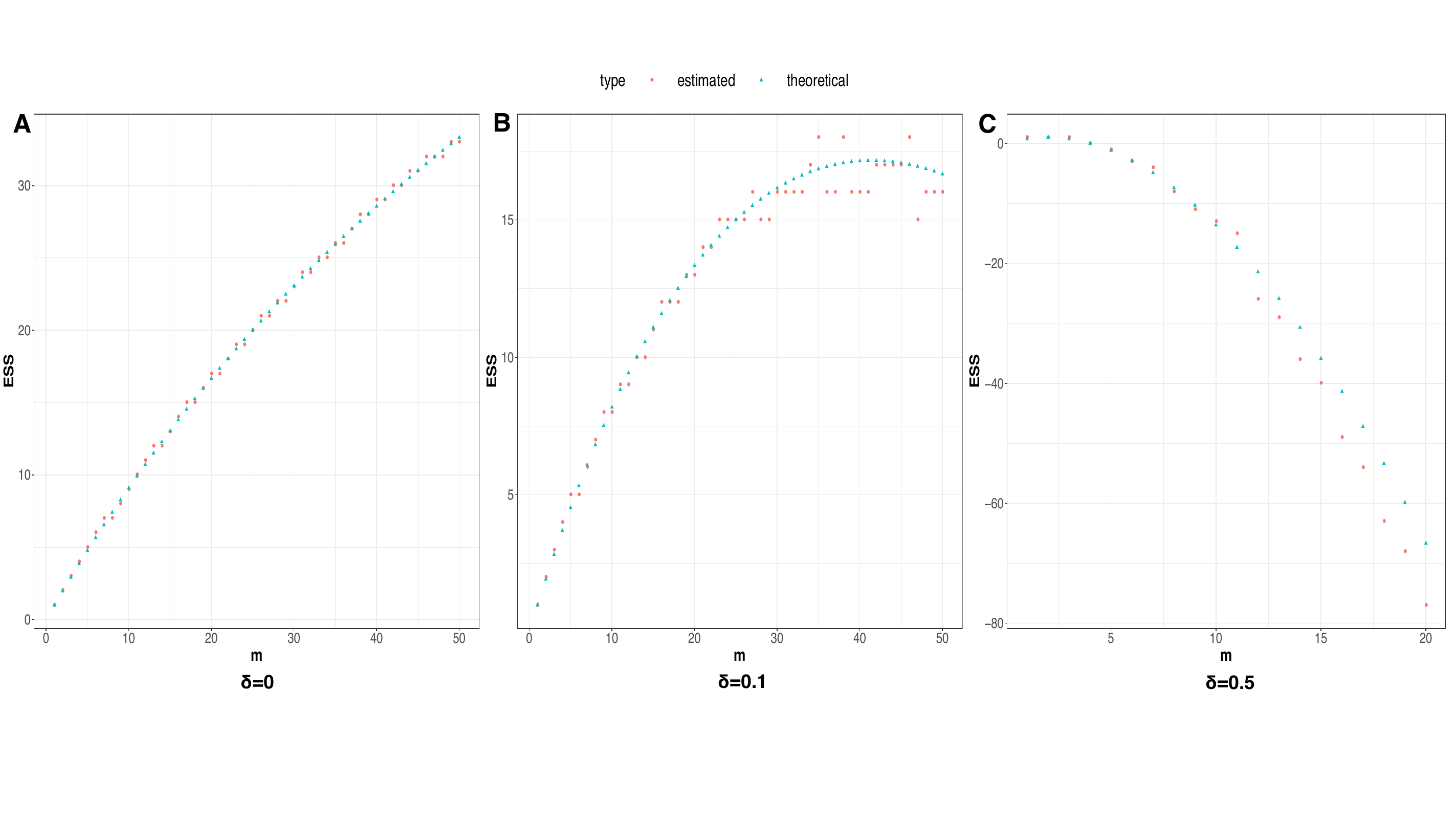}
	\caption{\textit{The estimated ESS versus $m$ under three scenarions: $\delta=0$, 0.1 and 0.5. The red dots are the estimated ESS, and the blue dots are the theoretical ESS.}}   
	\label{fig2}
\end{figure}

As a numerical illustration, the estimation procedure is conducted under three different scenarios: ($\textrm{i}$) no deviation; ($\textrm{ii}$) deviation with $\delta=0.1$; and ($\textrm{iii}$) deviation with $\delta=0.5$. First, the prior intensity level is fixed by letting $m=20$. The trend of distance $D(n,\tilde{n})$ across various values of ESS in the three scenarios is shown in Figure \ref{fig1}. It is evident that a larger deviation level pulls the ESS to a smaller value. Second, for the three scenarios, different prior intensity levels are explored by enumerating all integers between 1 and 50 for the value of $m$. The estimated ESS is also compared with its explicit form. As depicted in Figure \ref{fig2}, the increasing or decreasing tendency of the ESS curve is contingent on the deviation level. When the extent of the prior--likelihood disconcordance is small, a larger $m$ results in a larger ESS. Conversely, in cases with a substantial prior--likelihood disconcordance level, a larger prior intensity leads to a more detrimental prior.

\subsection{ESS of Beta prior in one-sample case}
\label{section1}

To estimate the ESS for a given Beta prior $\textrm{Beta}(a,b)$, we construct the hypothesis testing for parameters in its conjugate family---Bernoulli distribution. Scenarios where the prior is concordant or disconcordant with the likelihood are both considered.

Suppose that we have a set of Bernoulli samples, denoted as $\mathbb{X}_n=\{X_1,\ldots,X_{n}\}\sim \textrm{Bernoulli}(\theta)$. Consider the following hypothesis test,
$$H_0: \theta\leq \theta_0\quad \text{versus}\quad H_1: \theta>\theta_0.$$

In the frequentist framework, the $Z$-statistic is given by
$$Z_n^F=\frac{\bar{X}_n-\theta_0}{\sqrt{\theta_0(1-\theta_0)/n}},$$
where $\bar{X}_n=\sum_{i=1}^nX_i/n$ is the estimated proportion, and $S_n=\sum_{i=1}^nX_i\sim \textrm{Binomial}(n,\theta_0)$ is the binomial random variable. 

When switching into the Bayesian scheme, we impose a Beta prior on $\theta$, i.e., $\theta\sim \textrm{Beta}(a,b)$. Subsequently, the posterior distribution of $\theta$ is
$$\theta|S_n\sim \textrm{Beta}(a+S_n, b+n-S_n).$$
On the ground of the normal approximation for a beta distribution, we have
$$\textrm{Beta}(a+S_n, b+n-S_n)\approx N\left(\frac{a+S_n}{a+b+n}, \frac{(a+S_n)(b+n-S_n)}{(a+b+n)^2(a+b+n+1)}\right),$$
a $Z$-statistic in the Bayesian setting is naturally contructed as
$Z_n^B=\tilde{\mu}/\tilde{\sigma}$, where for simplicity, let
$$\tilde{\mu}=\frac{a+S_n}{a+b+n},\  \text{and} \ \ \tilde{\sigma}^2=\frac{(a+S_n)(b+n-S_n)}{(a+b+n)^2(a+b+n+1)}.$$
Following the rationale in the normal case, we also define two metrics,
$$U_n^B=E(Z_n^B)^2,\ \  U_{\tilde{n}}^F=E(Z_{\tilde{n}}^F)^2.$$
The distance $D(n,\tilde{n})=|U_n^B-U_{\tilde{n}}^F|$ is utilized to estimate the ESS. It is worth emphasizing that here $Z_{\tilde{n}}^F$ is defined as $Z_{\tilde{n}}^F=(\bar{X}_n-\theta_0)/\{\theta_0(1-\theta_0)/\tilde{n}\}^{1/2}$. In the one-sample case, without loss of generality, we consider the scenario where the underlying true distribution falls into the null hypothesis zone, i.e., $\theta_0=\theta$. Therefore, if $\tilde{n}^*=\argmin_{\tilde{n}}D(n,\tilde{n})$, the estimated ESS is $n-\tilde{n}^*$. 

\begin{figure}
	\centering         
	\includegraphics[width=0.95\textwidth]{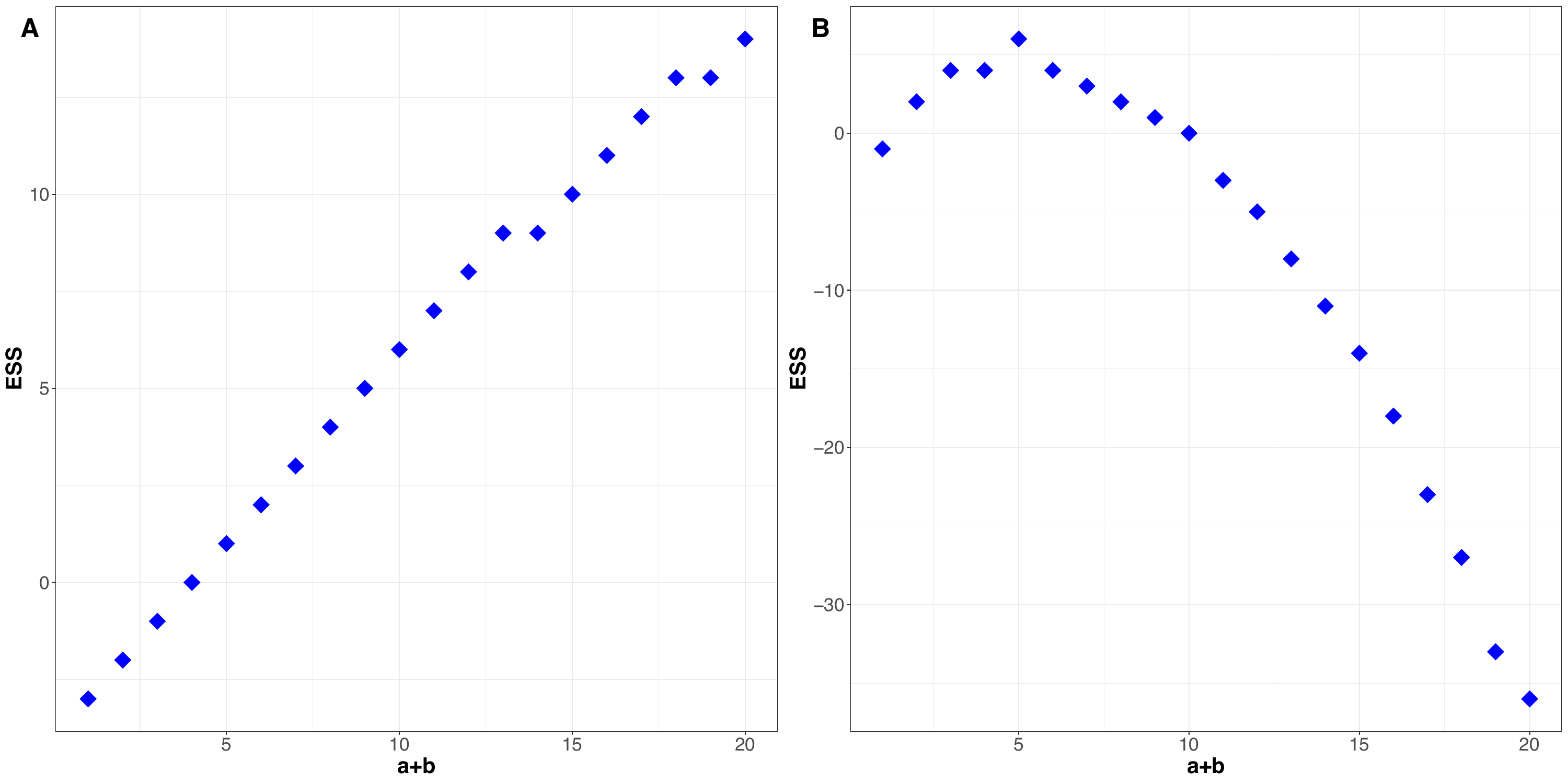}
	\caption{\textit{The ESS for a Beta prior under no deviation (A): $\theta_0=0.7$ and the prior mean equal to $\theta_0$ and deviation (B): $\theta_0=0.7$ and the prior mean equal to 0.5.}}   
	\label{fig4}
\end{figure}

In the subsequent numerical investigation, we focus on estimating the ESS under both deviation and no deviation cases while keeping $\theta_0=\theta=0.7$ fixed. First, with respect to the no deviation case, we set the prior mean to be equal to $\theta_0$, i.e., $\theta_0=a/(a+b)$, and enumerate all possible integers between 1 and 20 for the value of $a+b$. As shown in Figure \ref{fig4} (A), in this case, the ESS exhibits a linearly increasing trend with $a+b$, with the slope nearly equal to 1. This observation aligns with the well-established result that the ESS of $\textrm{Beta}(a,b)$ is equal to $a+b$. Second, we intentionally introduce deviation by setting the prior mean to deviate from the true value, specifically, we let $a/(a+b)=0.5$ and explore various values of $a+b$ from 1 to 20. Figure \ref{fig4} (B) shows that in the presence of deviation, the behavior of ESS is contingent on the strength of the prior, represented by $a+b$. When the deviation level is fixed (0.5 compared to 0.7), a larger prior strength leads to a more detrimental ESS. This finding echoes the observations made in the normal case.

\subsection{ESS of Beta prior in two-sample case}\label{two_sample_beta}

Suppose we have two sets of Bernoulli samples, detoted as $\mathbb{X}_n=\{X_1,\ldots,X_{n}\}\sim \textrm{Bernoulli}(\theta_1)$ and $\mathbb{Y}_n=\{Y_1,\ldots,Y_{n}\}\sim \textrm{Bernoulli}(\theta_2)$.  Consider the following hypothesis test,

$$H_0: \theta_1\leq \theta_2\quad \text{versus}\quad H_1: \theta_1>\theta_2.$$

In the frequentist paradigm, the $Z$-statistic is given by

$$Z_n^F=\frac{\bar{X}_n-\bar{Y}_n}{[\{{\bar{X}_n(1-\bar{X}_n)+\bar{Y}_n(1-\bar{Y}_n)}\}/n]^{1/2}},$$
where $\bar{X}_n=\sum_{i=1}^nX_i/n$ and $\bar{Y}_n=\sum_{i=1}^nY_i/n$ are the estimated proportions. Denote $S_n^X=\sum_{i=1}^nX_i\sim \textrm{Binomial}(n,\theta_1)$ and $S_n^Y=\sum_{i=1}^nY_i\sim \textrm{Binomial}(n,\theta_2)$. 

In the Bayesian paradigm, conjugate Beta priors are assigned for $\theta_1$ and $\theta_2$ respectively, i.e., $\theta_1\sim \textrm{Beta}(a_1,b_1)$ and $\theta_2\sim \textrm{Beta}(a_2,b_2)$. The posterior distributions of $\theta_1$ and $\theta_2$ are given by
$$\theta_1|S_n^X\sim \textrm{Beta}(a_1+S_n^X, b_1+n-S_n^X),\ \ \theta_2|S_n^Y\sim \textrm{Beta}(a_2+S_n^Y, b_2+n-S_n^Y).$$

Similar to the one-sample case, we introduce the notation,
$$\tilde{\mu}_X=\frac{a_1+S_n^X}{a_1+b_1+n},\ \ \tilde{\sigma}_X^2=\frac{(a_1+S_n^X)(b_1+n-S_n^X)}{(a_1+b_1+n)^2(a_1+b_1+n+1)},$$
$$\tilde{\mu}_Y=\frac{a_2+S_n^Y}{a_2+b_2+n},\ \ \tilde{\sigma}_Y^2=\frac{(a_2+S_n^Y)(b_2+n-S_n^Y)}{(a_2+b_2+n)^2(a_2+b_2+n+1)}.$$
The posterior distributions for $\theta_1$ and $\theta_2$ can be approximated by $N(\tilde{\mu}_X,\tilde{\sigma}_X^2)$ and $N(\tilde{\mu}_Y,\tilde{\sigma}_Y^2)$ respectively. Therefore, the corresponding $Z$-statistic can be constructed as $Z_n^B=\tilde{\mu}/\tilde{\sigma}$ with $\tilde{\mu}=\tilde{\mu}_X-\tilde{\mu}_Y$ and $\tilde{\sigma}^2=\tilde{\sigma}_X^2+\tilde{\sigma}_Y^2$. We again adopt the metric $D(n,\tilde{n})=|U_n^B-U_{\tilde{n}}^F|=|E(Z_n^B)^2-E(Z_{\tilde{n}}^F)^2|$ to evaluate the underlying ESS. 

One pivotal point to emphasize is that we have now incorporated ESS in the context of two-sample hypothesis testing. Unlike one-sample hypothesis testing, where the focus lies on parameter estimation, the ESS in two-sample testing is rooted in indicating how much sample size supports or hinders us in making correct decisions. The decision revolves around parameter comparison, and it is the collaborative impact of two priors that elicits the single ESS.

We illustrate the ESS under two general scenarios, where the ground truth is set to (a) $\theta_1=\theta_2$ and (b) $\theta_1>\theta_2$. For each scenario, we also examine two situations where the prior is concordant or disconcordant with the likelihood. It is essential to note that the definition of ESS differs in these two scenarios. For scenario (a), the ESS is defines as $\textrm{ESS}=n-\tilde{n}^*$ with $\tilde{n}^*=\argmin_{\tilde{n}}D(n,\tilde{n})$. In contrast, for scenario (b), where the underlying true distribution falls into the alternative hypothesis zone, we categorize priors with larger $U_n^B$ compared to $U_n^F$ as ``good" priors, as they yield higher power to reject $H_0$. Conversely, priors with smaller $U_n^B$ are categorized as ``poor" priors. Therefore, the ESS is defined as $\textrm{ESS}=\tilde{n}^*-n$. 

In scenario (a), the ESS estimation procedure is illustrated by fixing $\theta_1=\theta_2=0.4$. Priors supporting $\theta_1=\theta_2$ are considered as no-deviation priors. Given that two priors are to be specified, we categorize the no-deviation situation into several cases:
\begin{enumerate}[(i)]
	\item The same prior for $\theta_1$ and $\theta_2$, and both priors have no deviation from the truth;
	\item The same prior for $\theta_1$ and $\theta_2$, and both priors have the same deviation level from the truth;
	\item Different priors for $\theta_1$ and $\theta_2$, and the priors have no deviation from the truth but exhibit different prior strengths. 
\end{enumerate}

\begin{table}[H]
	\caption{\textit{The ESS of Beta prior under two-sample binomial data, with the true $\theta_1$ and $\theta_2$ equal to 0.4.  Priors with deviation or no deviation from the truth are considered.}} 
	\label{table:two-sample-binomial-0.4}
	\small
	\centering
	\begin{tabular}{cllr}
		\toprule
		Deviation & Prior of $\theta_1$ & Prior of $\theta_2$ & ESS \\ 
		\midrule
		No & $\textrm{Beta}(4,6)$ & $\textrm{Beta}(4,6)$ & 10 \\
		No & $\textrm{Beta}(2,3)$ & $\textrm{Beta}(2,3)$ & 5 \\
		No & $\textrm{Beta}(0.4,0.6)$ & $\textrm{Beta}(0.4,0.6)$ & 1 \\
		No & $\textrm{Beta}(1,9)$ & $\textrm{Beta}(1,9)$ & 7  \\
		No & $\textrm{Beta}(2,8)$ & $\textrm{Beta}(2,8)$ & 8   \\
		No & $\textrm{Beta}(3,7)$ & $\textrm{Beta}(3,7)$ & 9  \\
		No & $\textrm{Beta}(5,5)$ & $\textrm{Beta}(5,5)$ & 10  \\
		No & $\textrm{Beta}(6,4)$ & $\textrm{Beta}(6,4)$ & 11  \\
		No & $\textrm{Beta}(7,3)$ & $\textrm{Beta}(7,3)$ & 11  \\
		No & $\textrm{Beta}(8,2)$ & $\textrm{Beta}(8,2)$ & 12  \\
		No & $\textrm{Beta}(9,1)$ & $\textrm{Beta}(9,1)$ & 12  \\
		No & $\textrm{Beta}(4,6)$ & $\textrm{Beta}(2,3)$ & 7 \\
		No & $\textrm{Beta}(4,6)$ & $\textrm{Beta}(0.4,0.6)$ & 6  \\
		No & $\textrm{Beta}(4,6)$ & $\textrm{Beta}(0.04,0.06)$ & 3  \\	
		Yes & $\textrm{Beta}(4,6)$ & $\textrm{Beta}(2,8)$ & 0   \\
		Yes & $\textrm{Beta}(4,6)$ & $\textrm{Beta}(1,9)$ & $-8$  \\
		Yes & $\textrm{Beta}(4,6)$ & $\textrm{Beta}(5,5)$ & 8    \\
		Yes & $\textrm{Beta}(4,6)$ & $\textrm{Beta}(6,4)$ & 3   \\
		Yes & $\textrm{Beta}(4,6)$ & $\textrm{Beta}(7,3)$ & $-6$  \\
		Yes & $\textrm{Beta}(4,6)$ & $\textrm{Beta}(8,2)$ & $-20$  \\
		Yes & $\textrm{Beta}(4,6)$ & $\textrm{Beta}(9,1)$ & $-35$  \\
		\bottomrule
	\end{tabular}
\end{table}

As shown in Table \ref{table:two-sample-binomial-0.4}, with respect to case ($\textrm{i}$), the estimated ESS for two Beta priors exhibits remarkable similarity to that of one-sample Beta prior. For instance, the ESS for two distributions of $\textrm{Beta}(4,6)$, $\textrm{Beta}(2,3)$ and $\textrm{Beta}(0.4,0.6)$ are estimated to be 10, 5 and 1, respectively.
Regarding case ($\textrm{ii}$), the ESS possesses an intriguing behavior. For example, although $\textrm{Beta}(9,1)$ deviates from the true location 0.4, assigning $\textrm{Beta}(9,1)$ for both $\theta_1$ and $\theta_2$ results in an ESS of 12, which surpasses the ESS of 10 obtained with $\textrm{Beta}(4,6)$ in the one-sample no-deviation scenario. Therefore, the two $\textrm{Beta}(9,1)$ priors favor in making the correct decision to accept $H_0$. This blessing phenomenon originates from the fact that within the beta--binomial conjugate family, if the prior mean significantly deviates from the true mean of the binomial distribution, the resulting variance of the posterior beta distribution is inflated, leading to a reduced value of  $U_n^B$. Consequently, the smaller $U_n^B$ instills higher confidence in claiming that there is no difference between the two sample means, showcasing the blessing performance of the ESS. However, it is important to note that, the increasing tendency of ESS with ``$a$" in $\textrm{Beta}(a,b)$ suggests that $U_n^B$ is not solely determined by the distance between the means of prior and the data. 
In case ($\textrm{iii}$), where one prior is kept intact while the strength of the other prior varies, the ESS exhibits a clear increasing pattern with the strength of the other prior.

With respect to the deviation situation in scenario (a), it is evident that deviation conceals the inherent strengths of the two priors $\textrm{Beta}(a_1,b_1)$ and $\textrm{Beta}(a_2,b_2)$, leading to the ESS lower than either $a_1+b_1$ or $a_2+b_2$. When we have a poor guess of the two-group difference, the ESS may be substantially detrimental.

For scenario (b), we fix $\theta_1=0.7$ and $\theta_2=0.2$. The ESS is first evaluated in the situation where both priors have a perfect guess for the locations of $\theta_1$ and $\theta_2$. As shown in Table \ref{table:two-sample-binomial-0.7-0.2}, when the strengths of the two priors are equal, i.e., $a_1+b_1=a_2+b_2$ for priors $\textrm{Beta}(a_1,b_1)$ and $\textrm{Beta}(a_2,b_2)$, the ESS induced by two priors manifests a pattern similar to that observed in the one-sample no-deviation situation, it is equal to either $a_1+b_1$ or $a_2+b_2$.

Second, we investigate the deviation situation for scenario (b), where reasonable ESS estimation can be reached for both ``optimistic" and ``pessimistic" priors. When the difference between the priors exceeds the difference in the two-sample means, the ``optimistic" priors lead to a larger ESS, enabling us to discern the two groups with higher power. However, if the difference between the priors is smaller than the difference in the two-sample means, the priors may lead us astray in our decision-making process and result in a negative ESS. For instance, consider the $\textrm{Beta}(9,1)$ and $\textrm{Beta}(1,9)$ priors for $\theta_1$ and $\theta_2$ respectively. This prior setting suggests a substantial difference between the two groups, which is larger than the actual difference (0.8 versus 0.5). When the means of the two groups are indeed different, the ``optimistic" priors provide higher power in rejecting $H_0$ and lead to a large ESS. On the contrary, if we underestimate the difference, for example, by imposing the $\textrm{Beta}(6,4)$ and $\textrm{Beta}(5,5)$ priors for $\theta_1$ and $\theta_2$ respectively, the resulting ESS becomes $-11$. In this case, the ``pessimistic" prior hinders our correct decision in claiming the significant difference between the two groups. 

\begin{table}[H]
	\caption{\textit{The ESS of Beta prior under two-sample binomial data, with the true $\theta_1$ and $\theta_2$ are equal to 0.7 and 0.2, respectively.}} 
	\label{table:two-sample-binomial-0.7-0.2}
	\small
	\centering
	\begin{tabular}{cllr}
		\toprule
		Deviation & Prior of $\theta_1$ & Prior of $\theta_2$ & ESS \\ 
		\midrule
		No & $\textrm{Beta}(7,3)$ & $\textrm{Beta}(2,8)$ & 9 \\
		No & $\textrm{Beta}(3.5,1.5)$ & $\textrm{Beta}(1,4)$ & 5 \\
		No & $\textrm{Beta}(0.7,0.3)$ & $\textrm{Beta}(0.2,0.8)$ & 1 \\
		Yes & $\textrm{Beta}(7,3)$ & $\textrm{Beta}(7,3)$ & $-16$ \\
		Yes & $\textrm{Beta}(3.5,1.5)$ & $\textrm{Beta}(3.5,1.5)$ & $-9$  \\
		Yes & $\textrm{Beta}(0.7,0.3)$ & $\textrm{Beta}(0.7,0.3)$ & $-2$  \\
		Yes & $\textrm{Beta}(7,3)$ & $\textrm{Beta}(3,7)$ & 4  \\
		Yes & $\textrm{Beta}(7,3)$ & $\textrm{Beta}(4,6)$ & $-1$ \\
		Yes & $\textrm{Beta}(7,3)$ & $\textrm{Beta}(5,5)$ & $-6$  \\
		Yes & $\textrm{Beta}(7,3)$ & $\textrm{Beta}(1,9)$ & 15  \\
		Yes & $\textrm{Beta}(8,2)$ & $\textrm{Beta}(1,9)$ & 21  \\
		Yes & $\textrm{Beta}(9,1)$ & $\textrm{Beta}(1,9)$ & 27  \\
		Yes & $\textrm{Beta}(6,4)$ & $\textrm{Beta}(3,7)$ & $-1$  \\
		Yes & $\textrm{Beta}(6,4)$ & $\textrm{Beta}(4,6)$ & $-6$  \\
		Yes & $\textrm{Beta}(6,4)$ & $\textrm{Beta}(5,5)$ & $-11$ \\
		Yes & $\textrm{Beta}(5,5)$ & $\textrm{Beta}(5,5)$ & $-15$ \\
		\bottomrule
	\end{tabular}
\end{table}

\subsection{ESS in linear regression}\label{LR}

Consider two groups of normally distributed data $\mathbb{W}_{1,n_1}=\{W_{1,1},\ldots,W_{1,n_1}\}\sim N(\mu_1,\sigma^2)$ and $\mathbb{W}_{2,n_2}=\{W_{2,1},\ldots,W_{2,n_2}\}\sim N(\mu_2,\sigma^2)$. We aim to evaluate the equivalence of two-sample means by transferring it to a regression problem,
$$Y_i=\beta_0+\beta_1X_i+\epsilon,\ \ i=1,\ldots, n_1+n_2,$$
where $X_i=0$ and $Y_i=W_{1,i}$ for $1\leq i\leq n_1$, $X_i=1$ and $Y_i=W_{2,i-n_1}$ for $n_1+1\leq i\leq n_1+n_2$.

Without loss of generality, we assume $\mu_1=0$ and $\sigma^2$ is known. Our interest lies in the hypothesis test,
$$H_0:\mu_2\leq0\ \ \ \text{versus}\ \ \ H_1:\mu_2>0.$$
It is obvious that in the regression paradigm, the hypothesis testing problem can be translated into testing the slope,
$$H_0:\beta_1\leq0\ \ \ \text{versus}\ \ \ H_1:\beta_1>0.$$
From a frequentist perspective, the least square estimator $\hat{\beta}_1$ follows a normal distribution,
$$\hat{\beta}_1\sim N\left(\beta_1,\sigma^2\bigg{/}\sum\limits_{i=1}^{n_1+n_2}x_i^2\right).$$
The $Z$-score $Z_{n_1+n_2}^F=\hat{\beta}_1\left(\sum_{i=1}^{n_1+n_2}x_i^2\right)^{1/2}/\sigma$ is employed to make decisions regarding the acceptance or rejection of $H_0$.

\begin{figure}
	\centering    
	     
	\includegraphics[width=1\textwidth]{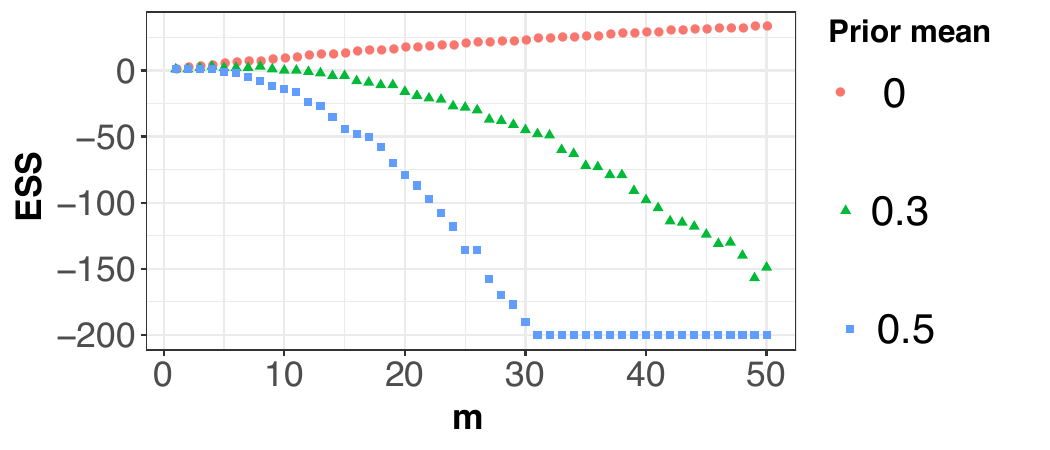}
	\caption{\textit{The ESS estimation under linear regression with no deviation (the prior mean is set as 0) and deviation (the prior mean is set as 0.3 and 0.5) cases. The underlying true situation is that the mean parameters of the two groups are both equal to 0. For all the three settings with different prior means, the prior intensity is varied by setting $m=1,\ldots,50$. The minimum value of ESS is truncated at $-200$.}}   
	\label{fig5}
\end{figure}	

In the Bayesian framework, assuming a normal prior for $\beta_1$, i.e., $\beta_1\sim N(\mu,\sigma_1^2)$, the corresponding posterior distribution for $\beta_1$ is given by
$$\beta_1\sim N\left(\frac{\sigma_1^2\left(\sum\limits_{i=1}^{n_1+n_2}x_iy_i\right)+\mu\sigma^2}{\sigma_1^2\left(\sum\limits_{i=1}^{n_1+n_2}x_i^2\right)+\sigma^2},\frac{\sigma^2\sigma_1^2}{\sigma_1^2\left(\sum\limits_{i=1}^{n_1+n_2}x_i^2\right)+\sigma^2}\right).$$
For simplicity, let $\tilde{\mu}$ and $\tilde{\sigma}^2$ denote the posterior mean and variance for $\beta_1$, respectively. The corresponding Bayesian $Z$-score is $Z_{n_1+n_2}^B=\tilde{\mu}/\tilde{\sigma}$.

\begin{table}[H]
	\caption{\textit{The ESS under the liner regression setting with unequal underlying mean parameters in the two groups, where $\mu_1=0$ and $\mu_2=0.3$. The prior mean for $\beta_1$ is set to be equal to 0 and 0.5, the prior variance of $\beta_1$ is varied by taking different values of $m$.}}
	\label{table:lr}
	\small
	\centering
	\begin{tabular}{ccc|ccr}
		\toprule
		Prior mean & Prior variance & ESS & Prior mean & Prior variance & ESS \\ 
		\midrule
		0 & 1 & $-1$ & 0.5 & 1 &2 \\
		0 & 1/3 & $-3$ & 0.5 & 1/3 & 7    \\
		0 & 1/6 & $-6$ & 0.5 & 1/6 & 14   \\
		0 & 1/9 & $-8$ & 0.5 & 1/9 & 21    \\
		0 & 1/12 &$ -11$ & 0.5 & 1/12 & 29    \\
		0 & 1/15 & $-13$ & 0.5 & 1/15 & 35    \\
		0 & 1/18 & $-15$ & 0.5 & 1/18 & 44   \\
		0 & 1/21 & $-17$ & 0.5 & 1/21 & 51    \\
		0 & 1/24 & $-19$ & 0.5 & 1/24 & 58    \\
		0 & 1/27 & $-21$ & 0.5 & 1/27 & 65   \\
		0 & 1/30 & $-23$ & 0.5 & 1/30 & 73    \\
		0 & 1/33 & $-25$ & 0.5 & 1/33 & 81   \\
		0 & 1/36 & $-26$ & 0.5 & 1/36 & 88    \\
		0 & 1/39 & $-28$ & 0.5 & 1/39 & 96    \\
		0 & 1/42 & $-30$ & 0.5 & 1/42 & 104   \\
		0 & 1/45 & $-31$ & 0.5 & 1/45 & 110    \\
		0 & 1/48 & $-32$ & 0.5 & 1/48 & 119   \\
		0 & 1/50 & $-33$ & 0.5 & 1/50 & 124    \\
		\bottomrule
	\end{tabular}
\end{table}

We proceed to utilize $D(n,\tilde{n})=|U_n^B-U_{\tilde{n}}^F|=|E(Z_n^B)^2-E(Z_{\tilde{n}}^F)^2|$ to estimate ESS within the context of linear regression. As in Section \ref{two_sample_beta}, the definition of ESS varies when the data align with the null hypothesis or the alternative hypothesis. When the data support $H_0$, $\textrm{ESS}=n-\tilde{n}^*$; when the data support $H_1$, $\textrm{ESS}=\tilde{n}^*-n$, where $\tilde{n}^*=\argmin_{\tilde{n}} D(n,\tilde{n})$.

In the numerical study, we delve into two distinct scenarios: (a) $\mu_2=0$ and (b) $\mu_2=0.3$. In the first scenario (a), no discernible mean disparity exists between the two groups. The ESS estimation is conducted for both deviation and no-deviation cases. In the no-deviation instance, we set $\mu_1=\mu_2=\mu=0$, $\sigma^2=1$, and modulate the prior's variance $\sigma_1^2=\sigma^2/m$ by varying the value of $m$. For the deviation scenario, we manipulate the prior mean $\mu$ and the prior variance $\sigma_1^2$ to gauge the ESS performance. As depicted in Figure \ref{fig5}, in the absence of deviation, the ESS displays an escalating trend with the increment of prior intensity (dictated by $\sigma_1^2$ or $m$). While in the presence of deviation, the ESS hinges on both the extent of deviation and the prior intensity. When these dual factors are substantial, the resultant negative ESS indicates a dilution in the statistical inference.

Transitioning to scenario (b), where a mean difference distinguishes the two groups, we set the prior mean $\mu$ for $\beta_1$ at either 0 or 0.5, corresponding to a ``pessimistic" and an ``optimistic" prior respectively. Here, $\mu=0.5$ bolsters the discrimination of the two groups with higher power, while $\mu=0$ prompts us to treat the two groups as homogeneous data. As presented in Table \ref{table:lr}, under the ``pessimistic" prior, $\mu=0$ hampers our ability to make accurate determinations in claiming the mean difference, as evidenced by the negative ESS. An elevated level of prior intensity exacerbates the negativity of ESS. Conversely, adopting the ``optimistic" prior where $\mu=0.5$ aligns the prior mean for $\beta_1$ to be larger than the true difference of 0.3. In this case, the computed ESS is positive and surpasses the value of $m$, indicating a power gain when testing the mean disparity between the two groups.

\begin{figure}
	\centering         
	\includegraphics[width=1\textwidth]{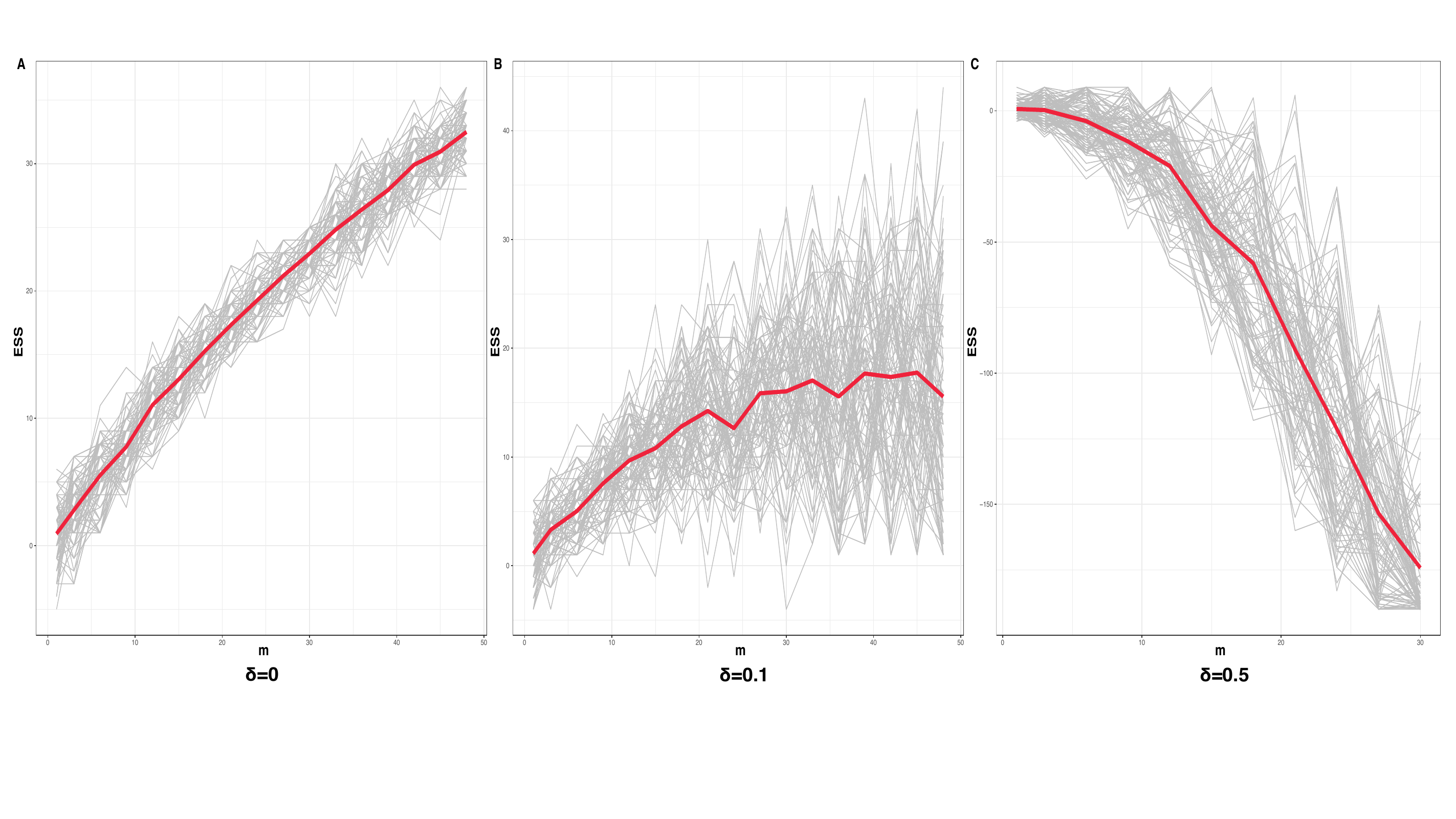}
	\caption{\textit{The ESS of normal prior versus $m$ with no deviation in (A) and deviation in (B) and (C) in simulation, where $m$ determines the prior variance. Each gray line is a replicate, and the red line is the averaged value of the 100 replicates.}}   
	\label{fig6}
\end{figure}

\section{Simulation}
In the simulation study, we conduct a comparative analysis between our $p$-value method and other ESS estimation methods proposed by \cite{reimherr2021prior} and \cite{morita2008determining} in the one-sample setting with normal prior and Beta prior. To emulate the case in real data analysis, we assume possession of a dataset with sample size 5,000. Employing bootstrap, 10,000 subsamples with replacement are created to construct the estimation of $D(n,\tilde{n})$. The Bayesian setting retains a fixed sample size of $n=100$. This procedure is replicated for 100 times to assess the robustness of the $p$-value ESS estimation procedure, with the final estimated ESS derived as the mean of these 100 replicates.

\begin{table}[H]
	\caption{\textit{The ESS of normal prior for three methods under no-deviation scenarios. The prior intensity is varied by changing different values of $m$. For each $m$, the $p$-value method and the method in \cite{reimherr2021prior} are replicated for 100 times, and the estimated ESS is taken as the average of the 100 replicates.}} 
	\label{table:simulation-normal-nodeviation}
	\small
	\centering
	\begin{tabular}{lccc}
		\toprule
		$m$ & $p$-value ESS & Remherr et al. & Morita et al.  \\ 
		\midrule
		1 & 0.93 & 1.74 & 1 \\
		3 & 2.79 & 4.51 & 3  \\
		6 & 5.52 & 9.17 & 6  \\
		9 & 7.76 & 13.70 & 9  \\
		12 & 11.04 & 18.31 & 12  \\
		15 & 13.05 & 22.84 & 15  \\
		18 & 15.27 & 27.89 & 18  \\
		21 & 17.33 & 32.46 & 21  \\
		24 & 19.26 & 37.47 & 24 \\
		27 & 21.19 & 42.11 & 27  \\
		30 & 22.93 & 46.80 & 30  \\
		\bottomrule
	\end{tabular}
\end{table}

In the normal prior case, the deviation level $\delta$ and prior intensity $m$ are varied for investigation. As shown in Figure \ref{fig6}, when no deviation is present ($\delta=0$), the estimated ESS exhibits an increasing trend as the prior intensity grows. For a moderate deviation level ($\delta=0.1$), the ESS curve displays a slower growth with increasing prior intensity compared with the case without deviation. However,
when the deviation level is more pronounced ($\delta=0.5$),  the ESS curve exhibits a sharp drop as shown in Figure \ref{fig6} (C). In the presence of deviation, a higher perturbation level is evident across the 100 replicates when the prior intensity is higher. On average, the ESS trend aligns with the pattern in Figure \ref{fig2}, where the data pool size is unrestricted and the distribution in the likelihood is presumed known. 

Compared with the methods in \cite{reimherr2021prior} and \cite{morita2008determining}, the results in Tables \ref{table:simulation-normal-nodeviation}--\ref{table:simulation-normal-deviation-0.5} underscore that both our $p$-value method and MSE method by \cite{reimherr2021prior} adaptly address the challenge of prior--likelihood disconcordance. In scenarios marked by significant deviation and elevated prior intensity, the prior tends to be detrimental to parameter inference. Additionally, our $p$-value ESS estimation method demonstrates higher sensitivity to the deviation level. As an example, in Table \ref{table:simulation-normal-deviation-0.5}, under equivalent deviation and prior intensity levels, the ESS derived from the $p$-value method exhibits more pronounced negativity compared with the MSE approach. In contrast, the method proposed by \cite{morita2008determining} yields an ESS determined solely by the prior intensity $m$, irrespective of the deviation level.

\begin{table}[H]
	\caption{\textit{The ESS of normal prior for three methods under the deviation scenarios, where the true mean in the likelihood is equal to 0, and the prior mean is set as 0.1. The prior intensity is varied by changing different values of $m$. For each $m$, the $p$-value method and the method in \cite{reimherr2021prior} are replicated for 100 times, and the estimated ESS is taken as the average of the 100 replicates.}} 
	\label{table:simulation-normal-deviation-0.1}
	\small
	\centering
	\begin{tabular}{lccc}
		\toprule
		$m$ & $p$-value ESS & Remherr et al. & Morita et al. \\ 
		\midrule
		1 & 1.12 & 2.49 & 1 \\
		3 & 3.31 & 5.59 & 3  \\
		6 & 5.03 & 10.09 & 6  \\
		9 & 7.57 & 14.17 & 9  \\
		12 & 9.68 & 18.55 & 12  \\
		15 & 10.81 & 22.72 & 15  \\
		18 & 12.82 & 26.98 & 18  \\
		21 & 14.24 & 30.85 & 21  \\
		24 & 12.65 & 34.59 & 24 \\
		27 & 15.86 & 38.53 & 27  \\
		30 & 16.04 & 42.44 & 30  \\
		\bottomrule
	\end{tabular}
\end{table}

\begin{table}[H]
	\caption{\textit{The ESS of normal prior for three methods under the deviation scenarios, where the true mean in the likelihood is equal to 0, and the prior mean is set as 0.5. The prior intensity is varied by changing different values of $m$. For each $m$, the $p$-value method and the method in \cite{reimherr2021prior} are replicated for 100 times, and the estimated ESS is taken as the average of the 100 replicates.}} 
	\label{table:simulation-normal-deviation-0.5}
	\small
	\centering
	\begin{tabular}{lccc}
		\toprule
		$m$ & $p$-value ESS & Remherr et al. & Morita et al.  \\ 
		\midrule
		1 & 0.71 & 1.33 & 1 \\
		3 & 0.30 & 3.31 & 3  \\
		6 & $-3.9$1 & 4.49 & 6  \\
		9 & $-11.67$ & 3.74 & 9  \\
		12 & $-21.00$ & 1.00 & 12  \\
		15 & $-43.76$ & $-2.48$ & 15  \\
		18 & $-58.12$ & $-7.14$ & 18  \\
		21 & $-91.01$ & $-11.28$ & 21  \\
		24 &$ -121.19$ & $-26.63$ & 24 \\
		27 & $-153.46$ & $-21.03$ & 27  \\
		30 & $-174.23$ & $-25.77$ & 30  \\
		\bottomrule
	\end{tabular}
\end{table}

\begin{figure}
	\centering         
	\includegraphics[width=1\textwidth]{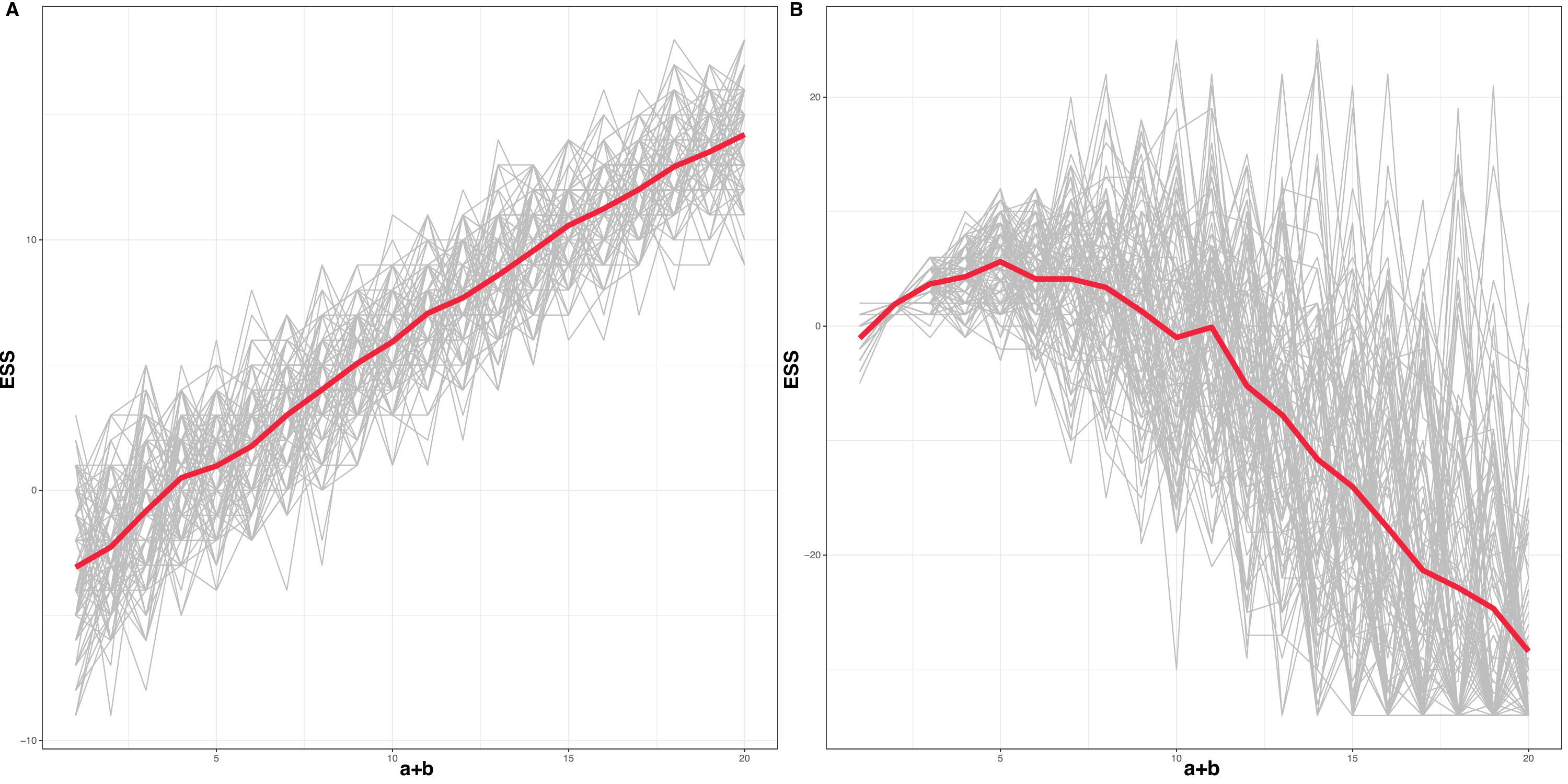}
	\caption{\textit{The ESS of Beta prior under no deviation (A) and deviation (B) scenarions with $\theta_0$ equal to 0.7. The prior mean is set as 0.7 (A) or 0.5 (B), and $a+b$ takes an integer value from 1 to 20. Each gray line is a replicate, the red line is the average of the 100 replicates.}}   
	\label{fig20}
\end{figure}

We proceed with a simulation study with respect to the Beta prior. As depicted in Figure \ref{fig20}, the estimated ESS curve closely mirrors that of Figure \ref{fig4}, wherein we assume a known distribution in both the likelihood and prior. In the absence of deviation, the perturbation levels across replicates display relatively consistent behavior across various $a+b$ values. However, in scenarios involving deviation, higher values of $a+b$ induce increased perturbation levels, suggesting that our $p$-value ESS estimation method exhibits heightened robustness when $a+b$ is small. Nevertheless, on average, a higher level of deviation leads to more negative ESS values for larger $a+b$ values.

In relation to the methods proposed by \cite{reimherr2021prior} and \cite{morita2008determining} within the Beta prior framework, their performances align with those observed in the normal prior context. Both the $p$-value method and the MSE method can effectively resolve the prior--likelihood disconcordant case. The method in \cite{morita2008determining} yields equivalent ESS estimation across varying deviation levels. All three methods concur that in the absence of deviation, the ESS demonstrates an increasing trend with $a+b$ in $\textrm{Beta}(a,b)$. In situations where deviation is present, both the $p$-value and MSE methods can produce very negative values for large $a+b$.

\begin{table}[H]
	\caption{\textit{The ESS of Beta prior for three methods under no-deviation scenarios, where the mean parameter $\theta$ in the likelihood is equal to 0.7. When imposing a Beta prior on $\theta$, the prior mean is equal to 0.7, $a+b$ in $\textrm{Beta}(a,b)$ is varied from 1 to 20. For each value of $a+b$, the $p$-value method and the method in \cite{reimherr2021prior} are replicated for 100 times, and the estimated ESS is taken as the average of the 100 replicates.}} 
	\label{table:simulation-beta-nodeviation}
	\small
	\centering
	\begin{tabular}{lccc}
		\toprule
		$a+b$ & $p$-value ESS & Remherr et al. & Morita et al.  \\ 
		\midrule
		1 & $-2.91$ & 1.81 & 1 \\
		2 & $-2.03$ & 2.95 & 2  \\
		3 & $-0.85$ & 4.77 & 3  \\
		4 & 0.33 & 6.08 & 4  \\
		5 & 1.31 & 7.79 & 5  \\
		6 & 2.30 & 9.13 & 6  \\
		7 & 3.36 & 10.56 & 7  \\
		8 & 4.18 & 12.12 & 8  \\
		9 & 5.26 & 13.76 & 9  \\
		10 & 5.85 & 15.21 & 10  \\
		11 & 6.75 & 16.61 & 11  \\
		12 & 7.83 & 18.33 & 12  \\
		13 & 8.93 & 20.07 & 13  \\
		14 & 9.62 & 21.64 & 14  \\
		15 & 10.39 & 23.07 & 15  \\
		16 & 11.22 & 24.52 & 16  \\
		17 & 12.06 & 26.00 & 17  \\
		18 & 12.77 & 27.66 & 18  \\
		19 & 13.67 & 29.04 & 19  \\
		20 & 14.12 & 30.80 & 20  \\
		\bottomrule
	\end{tabular}
\end{table}

\section{Real Data Application}
We apply the $p$-value ESS estimation method to an expression quantitative trait loci (eQTL) analysis, where eQTL are single-nucleotide polymorphisms (SNPs) that partly explain the variation in gene expression \citep{nica2013expression}. Our study leverages the Religious Orders Study and Rush Memory Aging Project (ROS/MAP) dataset for downstream analysis \citep{ng2017xqtl, bennett2018religious}, originally collected for aging and dementia research. After data preprocessing, there are 576 samples with both genotype and gene expression profiles for subsequent eQTL analysis. The eQTL is conventionally detected by regressing genotype information for each SNP against gene expression information,
$$\bm{Y}=\bm{X}\beta+\bm{\epsilon},$$
where $\bm{Y}$ represents the normalized expression vector, $\bm{X}$ is the genotype vector and $\bm{\epsilon}$ is the error.

\begin{table}[H]
	\caption{\textit{The ESS of Beta prior for three methods under the deviation scenarios, where the mean parameter $\theta$ in the likelihood is equal to 0.7. When imposing a Beta prior on $\theta$, the prior mean is equal to 0.5, $a+b$ in $\textrm{Beta}(a,b)$ is varied from 1 to 20. For each value of $a+b$, the $p$-value ESS estimation method and the method in \cite{reimherr2021prior} are replicated for 100 times, and the estimated ESS is taken as the average of the 100 replicates.}} 
	\label{table:simulation-beta-deviation}
	\small
	\centering
	\begin{tabular}{lccc}
		\toprule
		$a+b$ & $p$-value ESS & Remherr et al. & Morita et al. \\ 
		\midrule
		1 & $-1.19$ & 1.10 & 1 \\
		2 & 1.93 & 2.29 & 2  \\
		3 & 3.74 & 3.06 & 3  \\
		4 & 4.03 & 3.55 & 4  \\
		5 & 5.34 & 4.06 & 5  \\
		6 & 3.94 & 4.07 & 6  \\
		7 & 4.04 & 4.57 & 7  \\
		8 & 1.92 & 4.76 & 8  \\
		9 & 1.64 & 4.09 & 9  \\
		10 & $-0.02$ & 3.88 & 10  \\
		11 & $-2.53$ & 2.99 & 11  \\
		12 & $-5.54$ & 2.40 & 12  \\
		13 & $-10.08$ & 2.22 & 13  \\
		14 & $-9.76$ & 0.79 & 14  \\
		15 & $-14.27$ & 0.55 & 15  \\
		16 & $-16.5$5 & $-0.41$ & 16  \\
		17 & $-21.28$ & $-2.06$ & 17  \\
		18 & $-24.75$ & $-2.92$ & 18  \\
		19 & $-24.84$ & $-4.38$ & 19  \\
		20 & $-25.21$ & $-6.01$ & 20  \\
		\bottomrule
	\end{tabular}
\end{table}

\begin{table}[H]
	\caption{\textit{The ESS and the average of $|Z_n^B|$, denoted by $\overline{|Z_n^B|}$, under different normal priors on $\beta$, where $\beta$ is the effect size that SNP rs4813620 imposes on the expression of gene TRIB3.}}
	\label{table:realdata1}
	\small
	\centering
	\begin{tabular}{lcccr}
		\toprule
		Prior & $N(0.04, 1/100)$ &  $N(0.06, 1/100)$ & $N(0.07, 1/100)$ & $N(0.08, 1/100)$\\ 
		\midrule
		ESS & 8 & 62 & 90 & 119  \\
		$\overline{|Z_n^B|}$ & 1.821 & 1.910 & 1.953 & 1.998  \\
		\bottomrule
	\end{tabular}
\end{table}

In this study, our focus lies on exploring the eQTLs of genes APOE and TRIB3, which are widely recognized for their impact on Alzheimer's disease risk \citep{genin2011apoe, liu2018disease}.  Previous research has demonstrated that the SNPs rs4813620 and rs62109650 serve as eQTLs for genes TRIB3 and APOE, respectively \citep{liu2018disease, ramasamy2014genetic}. However, within the ROS/MAP cohort, partially due to the limited sample size, the estimated effect size of SNP rs4813620 on TRIB3 is $\hat{\beta}_1=0.0778$, with $p$-value equal to 0.0616; the estimated effect size of SNP rs62109650 on APOE is $\hat{\beta}_2=-0.0814$, with a corresponding $p$-value of 0.0505.

\begin{table}[H]
	\caption{\textit{The ESS and the average of $|Z_n^B|$, denoted by $\overline{|Z_n^B|}$, under different normal priors on $\beta$, where $\beta$ is the effect size that SNP rs62109650 imposes on the expression of gene APOE.}} 
	\label{table:realdata2}
	\small
	\centering
	\begin{tabular}{lcccr}
		\toprule
		Prior & $N(-0.04, 1/100)$ & $N(-0.06, 1/100)$ & $N(-0.07, 1/100)$ & $N(-0.08, 1/100)$\\  
		\midrule
		ESS & 3 & 55 & 81 & 109  \\
		$\overline{|Z_n^B|}$ & 1.895 &  1.983 & 2.033 & 2.077  \\
		\bottomrule
	\end{tabular}
\end{table}

Similar to the procedure outlined in Section \ref{LR}, we impose a normal prior on the coefficient $\beta$. The prior variance is fixed at $1/100$, a choice informed by the sample size in the Braineac database, where the sample size hovers around 100 for each tissue \citep{ramasamy2014genetic}. Given that the sample size of the complete data pool is 576, the sample size in the Bayesian setting is maintained to be $n=540$. The bootstrap is used to construct the metric $D(n,\tilde{n})$ for estimating the ESS. The results in Table \ref{table:realdata1} shed light on the eQTL analysis for gene TRIB3. When we possess more accurate estimation of $\beta$, the ESS converges towards 100, which is equivalent to the sample size of the ``historical" Braineac database. Specifically, when the prior mean is equal to 0.08, additional 119 samples are effectively incorporated into the eQTL detection procedure, which yields a posterior $Z$-statistics of 1.998, directing to clearer affirmation that rs4813620 serves as an eQTL for gene TRIB3. For the gene APOE, as depicted in Table \ref{table:realdata2}, similar results are obtained. It is noteworthy that even in scenarios where we underestimate the absolute value of the effect size, as exemplified by the imposition of a prior of $N(-0.06, 1/100)$ on $\beta$, there still is an augmentation of 55 samples and a $Z$-statistics of 1.983, significantly bolstering the identification of a significant eQTL.

\section{Discussion}
Despite being an age-old topic, great satisfaction has been achieved in witnessing continuous advancement in the ESS estimation field. From the very outset of focusing solely on the prior while neglecting the context in which the prior exists, the work by \cite{reimherr2021prior} has a milestone significance in inspiring new research on the ESS estimation, reminding us to focus on the underlying context in which the prior takes root.

By leveraging the connection between $p$-value and posterior probability under noninformative priors, we establish a novel approach to estimate the ESS, which allows for interpreting priors under different prior--likelihood deviation circumstances, thereby embracing both positive and negative ESS values. The proxy role of the $p$-value enables us to estimate the ESS without the burden of specification of a baseline prior. In essence, parameter estimation is the soil which ESS often finds its foundation in the realm of.
We address the parameter estimation by one-sample hypothesis testing. In addition to parameter estimation, we also place the ESS in the context of two-sample hypothesis testing, offering a powerful tool to evaluate the interplay of more-than-one priors, capturing the priors' optimistic or pessimistic roles on hypothesis testing results. We acknowledge that when dealing with multiple priors, our $p$-value ESS estimation method, within its current framework, yields a single ESS value that captures the interplay among the priors. To gain an insight into the distinct contributions and directions of influence exerted by individual priors on the resulting ESS, further decomposition strategies are imperative.

Obviously, the domain where prior distribution embeds in extends far beyond hypothesis testing or parameter inference. The ESS estimation within a specific domain necessitates specialized estimation methods, which deserves further explorations to harness the power of Bayesian statistical inference.

\clearpage
\bibliography{myReferences}
\end{document}